\documentclass[nohyper,11pt,letterpaper]{JHEP3}
\usepackage{amsfonts,amssymb}

\newcommand{\be}{\begin{equation}}
\newcommand{\ee}{\end{equation}}
\newcommand{\ben}{\begin{displaymath}}
\newcommand{\een}{\end{displaymath}}
\newcommand{\bea}{\begin{eqnarray}}
\newcommand{\eea}{\end{eqnarray}}
\newcommand{\bean}{\begin{eqnarray*}}
\newcommand{\eean}{\end{eqnarray*}}

\newcommand{\calk}{\mbox{${\cal K}$}}

\newcommand{\bbr}[1]{\mbox{${\mathbb R}^{#1}$}}
\newcommand{\bbz}[1]{\mbox{${\mathbb Z}_{#1}$}}

\newcommand{\ads}[1]{\mbox{${AdS}_{#1}$}}

\newcommand{\ie}{{\it i.e.}}

\newcommand{\beq}{\begin{equation}}
\newcommand{\eeq}{\end{equation}}
\newcommand{\beqr}{\begin{displaymath}}
\newcommand{\eeqr}{\end{displaymath}}
\newcommand{\beqa}{\begin{eqnarray}}
\newcommand{\eeqa}{\end{eqnarray}}
\newcommand{\beqar}{\begin{eqnarray*}}
\newcommand{\eeqar}{\end{eqnarray*}}
\newcommand{\cN}{{\cal N}}

\newcommand{\cB}{{\cal B}}

\newcommand{\half}{\ensuremath{\frac{1}{2}}}

\newcommand{\axidil}{\ensuremath{\chi+i\,e^{-\phi}}}
\newcommand{\sltz}{\ensuremath{SL(2,\bbz{})}}
\newcommand{\im}{\ensuremath{{\mathcal Im}}}
\newcommand{\bo}{\ensuremath{\bar{1}}}
\newcommand{\bt}{\ensuremath{\bar{2}}}

\newcommand{\bj}{\ensuremath{\bar{\jmath}}}

\newcommand{\bl}{\ensuremath{\bar{l}}}
\newcommand{\ba}{\ensuremath{\bar{a}}}
\newcommand{\bb}{\ensuremath{\bar{b}}}
\newcommand{\bc}{\ensuremath{\bar{c}}}
\newcommand{\bd}{\ensuremath{\bar{d}}}
\newcommand{\bz}{\ensuremath{\bar{z}}}
\newcommand{\bw}{\ensuremath{\bar{w}}}
\newcommand{\bzeta}{\ensuremath{\bar{\zeta}}}
\newcommand{\btau}{\ensuremath{\bar{\tau}}}
\newcommand{\bA}{\ensuremath{\bar{A}}}
\newcommand{\bB}{\ensuremath{\bar{B}}}
\newcommand{\bC}{\ensuremath{\bar{C}}}

\newcommand{\N}[1]{\ensuremath{\cN=#1}}
\newcommand{\tK}{\ensuremath{\tilde{K}}}
\newcommand{\Ai}{{\bf Ai}}

\newcommand{\besselK}{{\bf K}}
\newcommand{\teta}{\ensuremath{\tilde{\eta}}}
\newcommand{\bpartial}{\ensuremath{\bar{\partial}}}

\title{\LARGE Supergravity backgrounds corresponding to D7 branes wrapped on K\"ahler manifolds}

\author{Martin Kruczenski \\
        Department of Physics, Brandeis University \\
        Waltham, MA 02454.

E-mail: \email{martink@brandeis.edu}}

\abstract{ We consider supergravity solutions corresponding to D7 branes wrapped on K\"ahler manifolds with 
a $U(1)_R$ twist such that some supersymmetry is preserved. We find a class of $\frac{1}{4}$-BPS 
backgrounds where a $D7$-brane is wrapped on a $T^2$ torus with a metric of non-constant curvature.
Similarly to the flat $D7$-brane case, the solution has a singularity at finite radius. We
also discuss the case where the $D7$-brane is wrapped on a $4$-dimensional non-compact manifold. 
 The field theories on the D7 brane have \N{1} supersymmetry in 6 and 4 dimensions respectively.
}


\preprint{\tt{BRX TH-257} \\
          \tt{hep-th/0310225}  }

\begin{document}

\section{Introduction}

 The discovery of D-branes \cite{Polchinski:1995mt} led to an enormous progress in string theory.
At first, the interest was that they were the long sought string theory 
objects carrying RR-charge. However, it was soon understood that a perhaps
more important role they had was to allow the construction of many different
gauge theories as particular limits of string theory. Previously, that
was only possible by using the heterotic string.
Now, the world-volume theories on the D-branes were gauge theories with 
various matter contents (depending on the type of D-brane or D-brane 
intersection etc.) and many important properties of the gauge theories
were understood as simple geometrical or physical properties of 
D-branes\footnote{See \cite{Giveon:1998sr} for a review.}.

 For the purpose of this paper it is of particular interest the 
construction that makes use of D7 branes wrapped on K\"ahler manifolds 
\cite{Katz:1996th}. The idea is to wrap a D7 brane on a $2p$ dimensional 
complex manifold of  $U(p)=SU(p)\times U(1)$ holonomy (K\"ahler manifold) 
times $(8-2p)$-dimensional flat space. 
If the $U(1)$ part of the holonomy does not vanish, then there are no 
covariantly constant spinors and supersymmetry is completely broken.
However, the theory has an $U(1)_R$ R-symmetry under which the fermions
transform non-trivially. It is possible to gauge the $U(1)_R$ and introduce a 
fixed connection in such a way that it cancels the $U(1)$ part of the 
holonomy. 
 This is known as twisting the theory. The resulting theory preserves
one supercharge in $8-2p$ dimensions. Note that this is different from the case where the 
holonomy is $SU(p)$ (Calabi-Yau manifold) since there two
supercharges are preserved. They have opposite charges under the $U(1)$ but the
same under the $U(1)_R$ so, if the $U(1)$ is non-trivial only one of them 
can be preserved by twisting. 

 The supergravity background corresponding to such construction should have a
non-trivial metric, axion and dilaton. The axion and dilaton can be lumped 
together in the complex field
\beq
\tau= \chi + i e^{-\phi}.
\label{eq:taudef}
\eeq
 Type IIB theory is invariant under an $SL(2,\bbz{})$ duality 
which acts on $\tau$ as
\beq
\tau \rightarrow\frac{a\tau+b}{c\tau+d},\,\,\, a,b,c,d\in \bbz{}, \ \ ab-cd=1.
\label{sl2z}
\eeq
 It follows that $\tau$ does not need to be a globally well defined function, but
that in different coordinate patches one can have different functions as
long as the relation between them is as in (\ref{sl2z}). 

 As we review later in the paper, some amount of supersymmetry is preserved if we consider
the metric to be of the type
\beq
ds^2 = ds^2_{[1,9-2p]} + \partial_{a\bb} K(z^a,\bz^{\bb})\, dz^a d\bz^{\bb} , 
\label{metricans}
\eeq
where $ds^2_{[1,9-2p]}$ is a flat metric of signature $[1,9-2p]$. The 
$z^a$, $(a=1\ldots p)$ are complex coordinates and the K\"ahler potential $K(z^a,\bz^{\bb})$
has to satisfy
\beq
\det [ \partial_{a\bb} K(z^a,\bz^{\bb}) ] = \Omega(z^a)\bar{\Omega}(z^{\bb}) \frac{(\tau(z^a)-\btau(\bz^{\bb}))}{2i} ,
\label{maineq0}
\eeq
with $\Omega(z^a)$ and $\tau(z^a)$ arbitrary holomorphic functions except for the fact that
$\im (\tau) = e^{-\phi} >0$.

 One further point is that $\tau$ can be thought of as the modular
parameter of a flat torus $T^2$. By incorporating this torus explicitly, 
one can write down a 12-dimensional Ricci flat metric such that the K\"ahler manifold
and the torus give rise to a $p+1$ complex dimensional Calabi-Yau manifold.
This constructions is known as F-theory \cite{Vafa:1996xn} and from this point of view, the D7-branes 
are Calabi-Yau compactifications of F-theory. 

 In the rest of the paper we consider the cases $p=1,2$ and find non
trivial solutions to (\ref{maineq0}). The metric (\ref{metricans}) thus obtained
has curvature singularities in the same way as in the flat $D7$-brane ($p=0$) case.

\section{D7-brane solutions}
\label{sec:D7sol}

 D7-brane solutions were pioneered in \cite{Greene:1989ya} although in a different context. 
What was found there are string theory compactification on elliptically fibered
Calabi-Yau manifolds. The Calabi-Yau can be described as a certain (K\"ahler) base 
space $\cB$ over which a torus of modular parameter $\tau$ is fibered. By that, 
one means that $\tau$ is a given (holomorphic) function on the base space. The 
construction made explicit use of the fact that such function $\tau$ 
only needs to be defined up to modular transformations to construct supergravity solutions such 
that $\tau$ had non-trivial monodromies around the `core' of the solution. This is in fact 
a generalization of the more familiar construction of vortices in superconductors or cosmic strings. 

 It was realized in \cite{Gibbons:1995vg} that the same construction could be made in type
IIB theory since, due to the $\sltz$ invariance of type IIB, the field
$\tau=\axidil$ is also defined up to a modular transformation. The solution 
represents IIB theory compactified on $\cB$. 

 Afterwards, another D7-brane solution was found in \cite{Bergshoeff:1996ui} and described as a 
circularly symmetric D7-brane. Later, in \cite{Einhorn:2000ct} it was realized that they were both part
of the same construction. The idea was that one can find a solution of type IIB
supergravity with metric 
\beq
ds^2 = dx_{[1,7]}^2 + \Omega(z)\bar{\Omega}(\bar{z}) \tau_2 dz d\bar{z} ,
\label{D7metric}
\eeq
and with $\Omega$ and $\tau(z)=\tau_1+i\tau_2$ arbitrary\footnote{Actually there are some 
global restrictions if supersymmetry is to be preserved. See section 3.2 in \cite{Bergshoeff:2002mb} 
for a discussion. I am grateful to Ulf Gran for pointing out this to me.} holomorphic functions
of $z$.  This is a solution which locally preserves 16 real supercharges of IIB. 
Different choices of the functions functions lead to different solutions. 
Note that all these solutions have curvature singularities for certain values of $z$.

\section{Wrapped D7-branes}
\label{sec:WrappedD7}

In this paper we consider solutions describing wrapped D7-branes. These solutions 
will have only excited the axidilaton $\tau=\axidil$ and the metric. It is known that
when wrapping D-branes on curved manifolds, charges corresponding to lower dimensional branes
may arise. Here, we will consider that if that is the case, then corresponding
lower dimensional branes have been included to cancel the induced charge.

 Under these circumstances, the equations of motion (in Einstein frame) that we want to solve are \cite{Schwarz:qr}:
\beqa
R_{\mu\nu} &=& \frac{1}{2}\left(\partial_\mu \phi \partial_\nu \phi
                        + e^{2\phi} \partial_\mu \chi \partial_\nu\chi \right) ,\nonumber \\
\triangle \phi &=& e^{2\phi} g^{\mu\nu} \partial_\mu \chi \partial_\nu \chi \label{eq:eqomdil} ,\\
\triangle \chi  &=& -2g^{\mu\nu} \partial_\mu\phi\partial_\nu\chi \nonumber ,
\eeqa
where $\triangle$ denotes the covariant laplacian:
\beq
\triangle \phi = \frac{1}{\sqrt{-g}} \partial_\mu \left( \sqrt{-g} g^{\mu\nu} \partial_\nu \phi\right).
\eeq
 In terms of $\tau$ they can be written as
\beqa
R_{\mu\nu} &=& \frac{1}{4\tau_2^2}\left(\partial_\mu\tau\partial_\nu\btau+\partial_\nu\tau\partial_\mu\btau\right)\nonumber , \\
\triangle \tau  &=& \frac{2}{\tau-\btau} g^{\mu\nu} \partial_\mu\tau \partial_\nu \tau .
\label{taueq}
\eeqa
Notice, for later reference, that the equations are invariant under the rescaling
\beq
g_{\mu\nu} \rightarrow \lambda g_{\mu\nu} , \ \ \ \tau \rightarrow \sigma \tau ,
\label{rescale}
\eeq
for arbitrary constant $\lambda$, $\sigma$.

The solution we are seeking should be similar to (\ref{D7metric}) but with $dx^2_{[1,7]}$
replaced by $dx^2_{[1,7-2p]}+ ds^2_{[2p]}$ with $ds^2_{[2p]}$ the metric of the K\"ahler manifold
where the D7-brane is wrapped. This, in fact, cannot be the whole story since the metric 
$ds^2_{[2p]}$ can have parameters which depend on the transverse coordinates and we also
have to incorporate the twist that is used to preserve supersymmetry. Instead of searching
for an ansatz with this properties, it is simpler to start by analyzing the conditions under 
which the right amount of supersymmetry is preserved. We use an ansatz for the metric
\beq
ds^2 = dx_{[1,7-2p]}^2 + g_{a\bb} dz^a d\bar{z}^{\bb} ,
\label{eq:fullmetric}
\eeq
where $g_{a\bb}$ is a K\"ahler metric with K\"ahler potential $K$:
$g_{a\bb}=\partial_{a\bb}K$. This metric and the supersymmetry variations that
we compute below are in Einstein frame. Introducing a vielbein 
$e^A_a$, $\bar{e}^{\bA}_{\ba}$, the dilatino variation is given by 
\cite{Schwarz:qr}\footnote{We actually follow a simpler version which can be found in \cite{Bergshoeff:1996ui}.} 
\beq
\delta \lambda = -\frac{1}{2\tau_2} 
   \left( \partial_a\tau e^a_A \Gamma^A+\bpartial_{\ba} \tau e^{\ba}_{\bA} 
   \Gamma^{\bA}\right) \varepsilon^* .
\label{dilatino}
\eeq
If we now consider a spinor $\varepsilon$ annihilated by the $\Gamma^{\bA}$:
\beq 
\Gamma^{\bA} \varepsilon =0,\ \ \ \ \ \Gamma^{A} \varepsilon^*=0 ,
\label{eq:Ge}
\eeq
then we have that $\delta\lambda=0$ is satisfied if $\tau$ is holomorphic, 
namely $\bpartial_{\ba}\tau=0$. On the other hand, the gravitino variation is \cite{Schwarz:qr,Bergshoeff:1996ui}
\beqa
\delta \Psi_a &=& \partial_{a} \varepsilon + i \omega_{aA\bB} \Sigma^{A\bB}+ i \omega_{a\bA B} \Sigma^{\bA B} +
                  \frac{i}{4} \frac{\partial_a\tau_1}{\tau_2} \varepsilon  ,\nonumber \\
              &=& \partial_a \varepsilon 
                  + \frac{1}{4} \omega_{a A\bB} [ \Gamma^{A} ,\Gamma^{\bB} ] \varepsilon 
                  + \frac{1}{4} \omega_{a \bA B} [ \Gamma^{\bA} ,\Gamma^{B} ] \varepsilon
                  + \frac{i}{4} \frac{\partial_a \tau_1}{\tau_2} \varepsilon ,
\label{gravitino}
\eeqa
where we used that $\Sigma^{A\bB} = -\frac{i}{4}[\Gamma^A,\Gamma^{\bB}]$. Using the (anti)commutation 
properties of the Dirac matrices $\{\Gamma^A,\Gamma^{\bB}\} = \delta^{A\bB}$ and (\ref{eq:Ge}) we can 
recast this as
\beq
\delta \Psi_a = \partial_a \varepsilon - \frac{1}{2} \sum_A \omega_{a A\bA} \varepsilon
      + \frac{i}{4} \frac{\partial_a\tau_1}{\tau_2} \varepsilon  .
\label{dilatino2}
\eeq
Now, from Appendix B, we know that, for a K\"ahler manifold,
\beq
\sum_A \omega_{a A\bA} = -\partial_a \ln \det \bar{e}^{\bB}_{\bb} ,
\label{Uonecon}
\eeq
and so the gravitino variation vanishes if 
\beq
 \partial_a \ln \det \bar{e}^{\bB}_{\bb}
      = - \frac{i}{2} \frac{\partial_a \tau_1}{\tau_2}
      = \frac{1}{2} \partial_a \ln \tau_2 ,
\label{gravitino3}
\eeq
where we used $\partial_a\btau=0$ and the ansatz $\partial_a\varepsilon=0$. 
Eq.(\ref{gravitino3}) is solved if
\beq
\det \bar{e}^{\bA}_{\ba}= \bar{\Omega}(\bz^{\ba}) \sqrt{\tau_2} , 
\label{eeq}
\eeq
with an arbitrary holomorphic function $\Omega(z^a)$. From here we obtain
\beq
\det g_{a\bb} = \det e^{A}_{a} \det \bar{e}^{\bB}_{\bb} = \Omega(z^c) \bar{\Omega}(\bz^{\bc}) \tau_2 .
\label{maineq}
\eeq

We can always eliminate locally the function $\Omega$ by an appropriate change of coordinates.
Thus, we find that we need a K\"ahler manifold which admits a K\"ahler potential such that
\beq
\det g_{a\bb} = \im [\tau(z^c)] ,
\label{maineq2}
\eeq
in each coordinate patch and for a given holomerphic function $\tau(z^c)$. Notice that this has 
similarities with the Calabi-Yau case where
the condition is that $\det g = 1$ locally. In both cases, this implies one equation for
one indeterminate $K$ which in principle can be solved up to global obstructions. Instead of
further analyzing the mathematical properties of the metrics satisfying equation (\ref{maineq2}), 
in this paper we limit ourselves to look for examples. 

Before doing so, however, we should check that the equations of motion are also satisfied. 
It is easy to see that the axidilaton equation in (\ref{taueq}) is satisfied if $\tau$ is holomorphic.
In that case the equation for the metric reduces to
\beq
R_{a\bb} = \frac{1}{4} \frac{\partial_{a} \tau \bpartial_{\bb}\btau}{\tau_2^2}  .
\label{eqRtau}
\eeq
If we now use that
\beqa
R_{a\bb} &=& - \partial_{a\bb} \ln \det g_{c\bar{d}} , \\
\partial_{a\bb}\ln\tau_2 &=& -\frac{1}{4} \frac{\partial_{a} 
                               \tau \partial_{\bb}\btau}{\tau_2^2} , 
\label{Rln}
\eeqa
eq. (\ref{eqRtau}) becomes
\beq
- \partial_{a\bb} \ln \det g_{c\bar{d}} = -\partial_{a\bb}\ln\tau_2 ,
\eeq
which is satisfied in view of (\ref{maineq}).

 It was observed in \cite{Greene:1989ya} that, starting from a $2p$-dimensional metric that satisfies (\ref{maineq}) one
can construct a $2(p+1)$-dimensional Ricci flat metric. Introducing an extra complex coordinate 
$\zeta$  and defining a new K\"ahler potential as
\beq
\calk(z^a,\zeta,\bz^{\bb},\bzeta) = K(z^a,\bz^{\bb}) -\frac{(\zeta-\bzeta)^2}{\tau_2} ,
\label{cK}
\eeq
one gets a metric
\beq
ds^2 = \partial_{a\bb}K dz^a d\bz^{\bb} +\frac{1}{\tau_2} 
       \left| d\zeta-\frac{\zeta-\bzeta}{2i}\frac{\partial_a\tau dz^a}{\tau_2} \right|^2 .
\label{cKmetric}
\eeq
From here it follows that
\beq
\det g_{i\bj} = \frac{1}{\tau_2} \det \partial_{a\bb}K  ,
\label{gKeq}
\eeq
where $i=a,\zeta$ and $\bj=\bb,\bzeta$. Using (\ref{maineq}) this gives
\beq
\det g_{i\bj} = |\Omega(z^a)|^2 .
\label{CYeq}
\eeq
Since $R_{i\bj} = -\partial_{i\bj} \ln \det g_{k\bl}$ we get $R_{i\bj}=0$, namely the metric is Ricci flat. This means that 
the manifold is a Calabi-Yau manifold. Although the metric so constructed turns out to be singular at certain points, 
as explained in \cite{Greene:1989ya} it can be deformed to a non-singular one. In the context of type IIB theory in
which we are working this construction is called a compactification of F-theory \cite{Vafa:1996xn} on the (elliptically fibered) 
Calabi-Yau of K\"ahler potential $\calk$. The base of the fibration $\cB$ is the manifold we are studying.


\section{Complex dimension 1 (8 real supercharges)}
\label{d=2}

In this section we concentrate on the case $p=1$, namely, when wrapping the D7 brane on a 2-dimensional manifold. 
The resulting theory on the D7 brane is a 6-dimensional field theory with 8 supercharges. By further compactification 
on a flat $T^2$ torus we get an \N{2} four dimensional theory. 

  In view of the discussion of the previous section, we are looking for a function 
$K(z_1,z_2,\bz_1,\bz_2)$ such that 
\beq
\det \partial_{a\bb} K = \Omega(z^c) \bar{\Omega}(\bz^{\bc}) \tau_2 .
\label{Keq}
\eeq
Since $\tau$ is a holomorphic non-constant function we can locally use it as a coordinate and
identify $\tau=z_2$. The equation then becomes
\beq
K_{1\bo} K_{2\bt} - K_{1\bt} K_{2\bo} = |\Omega|^2 \frac{(z_2-\bz_2)}{2i} .
\label{Keq2}
\eeq
Although simple looking, this is a non-linear equation in partial derivatives which 
is very difficult to solve in full generality. The idea here is to find particular 
solutions using simplifying assumptions. The first such assumption is that $K$ 
depends only on the modulus of $z_1$. This implies that there is an $U(1)$ isometry
$z_1\rightarrow e^{i\theta} z_1$. Actually, for the purpose of the calculation it
is easier to do a coordinate transformation $w_1=\ln z_1$ and then $K$ will depend only 
on $w_1+\bw_1=\ln |z_1|^2$. Equivalently, we can consider that $K$ depends only on the real part
of $z_1$. We shall see that this assumption makes eq.(\ref{Keq2}) simpler to solve. 
A second and less necessary assumption is that $K$ depends only on $\tau_2$, the imaginary
part of $z_2$. Introducing real coordinates through
\beq
z_1 = y+i \phi, \ \ \ z_2 = \tau = \theta + i x ,
\label{realcoord}
\eeq
the assumption is that $K(z^a,\bz^{\ba}) = K(x,y)$. Thus, eq.(\ref{Keq2}) reduces to:
\beq
K_{yy} K_{xx} - K_{xy}^2 = x ,
\label{Keq3}
\eeq
where we did the final assumption that $\Omega=\mbox{cst.}=1/4$. Notice that in principle we could have 
had $|\Omega|= e^{\alpha x +\beta y}$. The assumption is that $\alpha=\beta=0$. The equation is still
a non-linear equation in partial derivatives but now depending only on two variables.
It can be solved by means of a Legendre transformation. Notice that if in the right hand side
we would have had 1 instead of $x$, then we would have been looking for a hyperK\"ahler 
manifold and the Legendre transform is a well-known method \cite{Lindstrom:rt}.

 The Legendre transform method starts by defining a new potential $\tK$ through
\beq
\tK(\eta,x) = K(y,x) - \eta y, \ \ \ \eta=K_y ,
\label{Leg}
\eeq
which implies
\beq
\tK_{\eta} = -y,\ \ \tK_x = K_x,\ \ K_{xx} = \tK_{xx}-\frac{\tK_{x\eta}^2}{\tK_{\eta\eta}} , \ \ \ 
K_{yy} = -\frac{1}{\tK_{\eta\eta}}, \ \ K_{xy} = -\frac{\tK_{x\eta}}{\tK_{\eta\eta}} .
\label{Leg2}
\eeq
Using these properties eq.(\ref{Keq3}) becomes
\beq
\tK_{xx} + x \tK_{\eta\eta} =0 ,
\label{eqLeg}
\eeq
which is now a linear differential equation. A particularly simple solution depending on one parameter ($a_0>0$) is 
\beq
\tK = a_0 (x^3-3\eta^2 ) ,
\label{simplesolution}
\eeq
which as we check below corresponds to a flat D7-brane. 
It is not much more difficult to find the most general solution (see Appendix for a derivation):
\beq
\tK(\eta,x) = a_0 (x^3-3\eta^2) - \frac{3^{\frac{1}{3}}}{2\pi^2} \Gamma\left(\frac{2}{3}\right)^3 
              \int_{-\infty}^{+\infty} 
              \frac{x}{\left[(\eta-\teta)^2+\frac{4}{9}x^3\right]^{\frac{5}{6}}} 
              h(\teta)d\teta  , 
\label{soldtwo}
\eeq
which depends on an arbitrary function $h(\eta)$ that, as we shall see, determines the
metric in the manifold over which the D7 is wrapped. The derivation of this expression
is given in the appendix. However it is easy to check that it solves eq.(\ref{eqLeg}) since
\beq
\left(\partial_{xx} + x \partial_{\eta\eta}\right) 
              \frac{x}{\left[\eta^2+\frac{4}{9}x^3\right]^{\frac{5}{6}}} = 0  .
\label{check}
\eeq
The coefficient in front of the integral in (\ref{soldtwo}) is chosen such that
\beq
\tK(x=0,\eta) = -3a_0\eta^2 - h(\eta)  .
\eeq
From here we can undo the Legendre transformation and get the K\"ahler potential
if we wished. However it is simpler to use $\eta$ as a coordinate and get the metric
directly. To see that, we start by writing the metric in the original coordinates
\beqa
ds^2 &=& K_{1\bo} dz_1 d\bz_1 + K_{1\bt} dz_1 d\bz_2 + 
       K_{\bo 2} d\bz_1 dz_2 + K_{2\bt} dz_2 d\bz_2 \\
     &=& \frac{1}{4K_{yy}} \left(K_{xx} K_{yy} -K_{xy}^2\right) dz_2 d\bz_2 + 
         \frac{1}{4}K_{yy} \left| dz_1 - i \frac{K_{xy}}{K_{yy}} dz_2 \right|^2 .
\label{metricdtwo}
\eeqa
Performing the Legendre transformation and using (\ref{realcoord}) we obtain 
\beq
ds^2 = \frac{1}{4} \tK_{xx}(dx^2+d\theta^2) -\frac{1}{4\tK_{\eta\eta}}\left[(dy+\tK_{x\eta}dx)^2
 + (d\phi-\tK_{x\eta}d\theta)^2\right] .
\eeq
We can change coordinates $y\rightarrow\eta$ noting that $dy+\tK_{x\eta}dx = -\tK_{\eta\eta}d\eta$ 
to get
\beq
ds^2 = -\frac{1}{4} \tK_{\eta\eta} \left[x(dx^2+d\theta^2)+d\eta^2\right] 
       - \frac{1}{4\tK_{\eta\eta}}\left(d\phi-\tK_{\eta x} d\theta\right)^2 ,
\label{metricdtwofinal}
\eeq
where we used eq.(\ref{eqLeg}) for $\tK$.
 One simple check is that the solution (\ref{simplesolution}) with $a_0=2/3$ gives the metric
\beq
ds^2 =  x(dx^2+d\theta^2)+ d\eta^2 + d\phi^2  ,
\eeq
which is the same as the metric (\ref{D7metric}) after choosing $\Omega = 1/z$, $\tau=i\ln z$ and $z=\exp(x+i\theta)$.

Another thing we need to ensure is that the metric is positive definite. As it stands, it is clear that it
would be positive definite as long as $\tK_{\eta\eta}<0$. To check that notice that (\ref{soldtwo}) 
can also be written as
\beq
\tK(\eta,x) =  a_0 (x^3-3\eta^2) - \frac{3^{\frac{1}{3}}}{2\pi^2} \Gamma\left(\frac{2}{3}\right)^3 
              \int_{-\infty}^{+\infty} 
              \frac{x}{\left[\teta^2+\frac{4}{9}x^3\right]^{\frac{5}{6}}} 
              h(\teta+\eta)d\teta  ,
\label{AKsol2}
\eeq
which shows that $\tK_{\eta\eta}$ is given by
\beq
\tK(\eta,x)_{\eta\eta} = -6 a_0 - \frac{3^{\frac{1}{3}}}{2\pi^2} \Gamma\left(\frac{2}{3}\right)^3 
              \int_{-\infty}^{+\infty} 
              \frac{x}{\left[\teta^2+\frac{4}{9}x^3\right]^{\frac{5}{6}}} 
              h_{\eta\eta}(\teta+\eta)d\teta  , 
\label{AKetaeta}
\eeq
which is not surprising since $\tK_{\eta\eta}$ satisfies the same equation (\ref{eqLeg}). The point,
however, is that we can rewrite this as 
\beq
\tK(\eta,x)_{\eta\eta} = - \frac{3^{\frac{1}{3}}}{2\pi^2} \Gamma\left(\frac{2}{3}\right)^3 
              \int_{-\infty}^{+\infty} 
              \frac{x}{\left[\teta^2+\frac{4}{9}x^3\right]^{\frac{5}{6}}} 
              (6 a_0 + h_{\eta\eta}(\teta+\eta))d\teta  , 
\label{AKetaeta2}
\eeq
so, as long as $\tK_{\eta\eta}$ is negative at $x=0$:
\beq
\tK_{\eta\eta}(x=0,\eta) = - (6a_0 + h_{\eta\eta}(\eta)) < 0 , 
\label{rightsign}
\eeq
then $\tK_{\eta\eta} <0$ for any value of $x$ since the integrand in (\ref{AKetaeta2}) is positive.
That means that as long as we choose $h(\eta)$ appropriately, namely satisfying (\ref{rightsign}),
the metric will be positive definite.

 As a further verification, we can start with the following ansatz
\beqa
ds^2 &=& 
   x\, f\, (dx^2+d\theta^2)+ f d\eta^2 + \frac{1}{f}\left(d\phi+B d\theta\right)^2 \nonumber\\
\phi &=& -\ln(Nx) \label{ansatzdtwo} , \\
\chi &=& N\theta  ,\nonumber \\
\tau &=& \chi + i e^{-\phi} = N(\theta+i x)  , \nonumber
\eeqa
and verify that the equations of motion (\ref{eq:eqomdil}) are satisfied if
\beq
f_{xx}+xf_{\eta\eta}=0 ,\ \ \ B_{\eta}=-f_x,\ \ \ B_x = xf_{\eta} .
\label{feq}
\eeq
Identifying $f=-\tK_{\eta\eta}$ and $B=\tK_{\eta x}$ we see that these last equations are implied by eq.(\ref{eqLeg}). 
In (\ref{ansatzdtwo}) we used the rescaling freedom we mention in eq.(\ref{rescale}) to scale away the $1/4$ in the
metric and to introduce a parameter $N$ that we associate with the number of $D7$-branes since now, when 
$\theta\rightarrow \theta+2\pi$ we get $\chi\rightarrow \chi+2\pi N$.

It is useful to compute the Ricci scalar for the metric (\ref{ansatzdtwo}) which is
\beq
R = \frac{1}{fx^3} .
\eeq
We see that, as long as $f>0$ the only singularity is at $x=0$. This singularity is the same as the one that appears for
the flat brane. Another fact that we can check using the curvature is that this solution is not just a change of coordinates
of the usual flat brane. The flat brane corresponds to constant $f$. We can write $f$ as
\beq
f = \frac{1}{x^3 R} = N^3 e^{3\phi} \frac{1}{R} .
\eeq
If $f$ is constant, we have
\beq
\nabla_\mu \frac{e^{3\phi}}{R} = 0 ,
\eeq
which is a coordinate independent statement. When $f$ is not constant the derivative does not vanish
and so the background is different. That is to say, it might be possible to redefine the coordinates so that
the metrics agree but then the functions $\phi$ will differ.

As a final check we can use the construction in eq.(\ref{cKmetric}) and find a Ricci flat metric
\beq
ds^2 =  f \left(x(dx^2+d\theta^2)+d\eta^2\right) + \frac{1}{f}\left(d\phi+B d\theta\right)^2  + 
 \frac{1}{x}\left(d\alpha -\frac{\beta}{x} d\theta\right)^2 +
 \frac{1}{x}\left(d\beta -\frac{\beta}{x} dx\right)^2 ,
\label{extended1}
\eeq
where we introduced real coordinates $\alpha$ and $\beta$ through
\beq
\zeta= \alpha+i\beta .
\eeq
A straightforward calculation shows that, given (\ref{feq}), the metric (\ref{extended1}) is Ricci flat.

The surfaces of constant dilaton and axion are the surfaces of constant $x=x_0$ and $\theta$. They are parameterized
by $\eta$ and $\phi$ with an induced metric given by
\beq
ds^2_{\mbox{ind.}} = f(x_0,\eta) d\eta^2 + \frac{1}{f(x_0,\eta)} d\phi^2 ,
\label{internalmetric}
\eeq
Since this is the metric induced on the surfaces of constant dilaton and axion, it characterizes the geometry around the D7 brane 
in a coordinate independent way and therefore should be related to the metric of the manifold where the D7 brane is wrapped. 
The precise relation is worked out after eq.(\ref{m3v4}) in a more general case\footnote{If we use the same procedure that 
leads to eq.(\ref{m3v4}) we get the same metric (\ref{internalmetric}) up to a factor $(1+\frac{B^2}{xf^2})^{-1}$. This factor 
is non-singular for $x\neq 0$ and therefore the topological properties of the manifold where the $D7$ brane is wrapped are 
the same as those of the manifolds with constant axidilaton that we analyze here}.  
By construction, the metric (\ref{internalmetric}) has an isometry $\phi\rightarrow\phi+\alpha$. The topology depends on 
the function $f(x_0,\eta)$. If $f>0$ for all values of $\eta$, $-\infty<\eta<\infty$ and 
$0<\lim_{\eta\rightarrow\pm\infty} f(x_0,\eta)<\infty$ then the manifold is non compact of topology $\bbr{1}\times S^1$. 
One can get a topology $S^2$ if the circle $\phi$ closes
which means that $f$ diverges at some point. For example the standard spherical round metric is obtained for $f=1/(1-\eta^2)$ 
as can be seen by replacing $\eta=\cos\Theta$. We see however that, if we use such a function $f$, 
the metric (\ref{ansatzdtwo}) will be singular at $\eta=\pm 1$ for any $x$ since $f$
also multiplies $dx^2+d\theta^2$. Therefore, the topology $S^2$ cannot be obtained from this ansatz. However we can still
obtain a compact manifold if $f$ is periodic in $\eta$. In that case we can periodically identify $\eta$ with the same period and
obtain a manifold whose topology is $T^2$ with a non-flat metric given by (\ref{internalmetric}).
 The solution in this case can be obtained by considering a generic periodic function 
\beq
h(\eta) = \sum _{n=1}^{\infty} a_n \cos(n \eta) ,
\eeq
where the $a_n$ are arbitrary coefficients restricted only by the fact that $h$, $h_{\eta}$ and $h_{\eta\eta}$ should be 
bounded continuous functions.
Replacing in (\ref{soldtwo}) gives
\beq
\tK = a_0(x^3-3\eta^2) - 3^{\frac{2}{3}} \Gamma\left(\frac{2}{3}\right) 
     \sum_{n=1}^{\infty} a_n \Ai\!\left(xn^{\frac{3}{2}}\right) \cos(n\eta) ,
\label{compact}
\eeq
where $\Ai$ denotes the Airy function that vanishes at infinity and $a_0$ should be chosen so as to ensure
that $\tK_{\eta\eta}(x=0,\eta)<0$ \ie\ $6a_0+h_{\eta\eta}>0$.   
The result can also be written using Bessel functions $\besselK_\nu$ since
\beq
\Ai(x) = \frac{1}{\pi} \sqrt{\frac{x}{3}} \besselK_{\frac{1}{3}}\!\left(\frac{2}{3}x^{\frac{3}{2}}\right) .
\label{Airy}
\eeq
As an example we can take
\beq
\tK = x^3-3\eta^2 - 3^{\frac{2}{3}} \Gamma\left(\frac{2}{3}\right) \Ai(x)
      \cos(\eta)  .
\label{compact1}
\eeq
Since the Airy function vanishes exponentially at infinity, we see that for $x\rightarrow\infty$ 
we have $\tK \simeq a_0(x^3-3\eta^2)$, that is the metric of the flat D7 brane. All the influence of the coefficients 
$a_{n\ge 1}$ disappears. This would suggest that large $x$ corresponds to the UV properties of the theory on the
D7 branes since at short distances the details of the metric are irrelevant. 
 

\section{Complex dimension 2 (4 real supercharges)}
\label{d=4}
In this section we study the case when $p=2$, that is when we wrap the D7 brane on a manifold of
dimension $d=4$.
In this case we were not able to linearize the equation as in the case $p=1$ ($d=2$). Again we have to
solve
\beq
\det \partial_{a\bb} K = \Omega(z_c) \bar{\Omega}(z_{\bc}) \tau_2 ,
\label{Keq4}
\eeq
where now $a,\bb=1\ldots 3$. We can always take $\tau=z_3$. If we now assume that $K$ is only
a function of $\tau_2$ we can simplify the equation. If we define $z_3=\theta+i x$ similarly as before, 
then eq.(\ref{Keq4}) can be written as
\beq
\det\left(
\begin{array}{cc}
K_{a\bb} & -\frac{1}{2i}K_{ax} \\
\frac{1}{2i} K_{\ba x} & \frac{1}{4} K_{xx}
\end{array}
\right) = \frac{1}{4}K_{xx}\det\left(K_{a\bb}-\frac{K_{ax}K_{\bb x}}{K_{xx}}\right) = \Omega(z_c) \bar{\Omega}(z_{\bc}) x ,
\label{Keq41}
\eeq
where now $a,\bb=1,2$ since we wrote the dependence in $z_3$ explicitly.
Now we can do a Legendre transformation from $x$ to $\eta$. Notice that this is different from what
we did before where the Legendre transformation was not with respect to $x$. In any case the calculations are similar.
We define
\beq
\tK = K -\eta x,\ \ \ \eta=K_x ,
\label{Legtransf}
\eeq
which results in 
\beq
K_a = \tK_a, \ \ \tK_\eta = -x ,\ \ K_{a\bb} = \tK_{a\bb}-\frac{\tK_{a\eta}\tK_{\bb\eta}}{\tK_{\eta\eta}}, \ \ 
K_{xx} = -\frac{1}{\tK_{\eta\eta}}, \ \ K_{ax} = -\frac{\tK_{a\eta}}{\tK_{\eta\eta}} .
\label{Legtransf2}
\eeq
Replacing in (\ref{Keq41}) we get
\beq
\det(\tK_{a\bb}) = 4 |\Omega|^2 \tK_\eta \tK_{\eta\eta} ,
\label{K3eq}
\eeq
where we are assuming that $\Omega$ is independent of $\tau$.
Since the equation did not linearize we can only search just for a particular solution rather than solving
it in general. In order to do that we use a factorized ansatz:  
\beq
\tK = \Phi(z^a,z^{\ba}) X(\eta) \ \ \ \Rightarrow \ \ \ X^2 \det(\Phi_{a\bb}) = 4 \Phi^2 X_\eta X_{\eta\eta} .
\label{factorize}
\eeq
With this ansatz we get two equations:
\beqa
X^2 &=& \pm 4 X_\eta X_{\eta\eta}  = \pm \frac{4}{3} \frac{d}{dX} X_{\eta}^3 \label{Xeq} \\
\det \Phi_{a\bb} &=& \pm \Phi^2  .
\label{phieq}
\eeqa
The first equation can be easily integrated giving
\beq
\eta = \int^{X} \frac{d\tilde{X}}{(A\pm\frac{\tilde{X}^3}{4})^{\frac{1}{3}}} ,
\label{Xint}
\eeq
where $A$ is an integration constant.
Before looking into the equation for $\Phi$ we find out what is the metric like under the assumptions
we have made so far. We have
\beq
ds^2 = K_{a\bb} dz^a d\bz^{\bb} - \frac{1}{2i} K_{ax} dz^a d\bz_3+ K_{\ba x} \frac{1}{2i} d\bz^{\ba} dz_3 
       + \frac{1}{4}K_{xx}dz_3d\bz_3  .
\label{m3v1}
\eeq
With the definition $z_3=\theta+ix$, and after some algebra we arrive at
\beq
ds^2 = \left(K_{a\bb} -\frac{K_{ax} K_{\bb x}}{K_{xx}} \right) dz^a d\bz^{\bb} + \frac{1}{4K_{xx}}\left(dK_x\right)^2
 + \frac{K_{xx}}{4}\left(d\theta+\frac{i}{K_{xx}}(K_{ax}dz^a-K_{\ba x}d\bz^{\ba})\right)^2 .
\label{m3v2}
\eeq
Now we transform coordinates from $x$ to $\eta$ and introduce the Legendre transform to get
\beq
ds^2 = \tK_{a\bb} dz^a d\bz^{\bb} - \frac{1}{4}\tK_{\eta\eta} d\eta^2 -\frac{1}{4\tK_{\eta\eta}} 
\left(d\theta + i (\tK_{a\eta}dz^a - \tK_{\ba\eta}d\bz^{\ba})\right)^2 .
\label{m3v3}
\eeq
The ansatz $\tK=\Phi X$ gives now
\beq
ds^2 = X\Phi_{a\bb} dz^a d\bz^{\bb} - \frac{1}{4}\Phi X_{\eta\eta} d\eta^2 -\frac{1}{4X_{\eta\eta}\Phi}
 \left(d\theta-iX_{\eta}(\Phi_a dz^a - \Phi_{\ba} d\bz^{\ba})\right)^2 .
\label{m3v4}
\eeq
We can consider $\eta$ as a radial variable. For fix $\eta$,  there is  a four dimensional 
K\"ahler manifold with K\"ahler potential $\Phi(z^a,z^{\bb})$ whose volume is proportional to $X(\eta)^4$ and
over which a circle parameterized by $\theta$ is fibered. The metric of the manifold where the $D7$ brane
is wrapped is then a K\"ahler metric with K\"ahler potential $\Phi$ and
the $S^1$ fibration is the supergravity equivalent of the $U(1)_R$ twist that we discussed at the beginning 
should be introduced in the theory on the brane to preserve supersymmetry. Another useful metric to consider 
is that of the surfaces of constant dilaton and axion that we discussed in the previous section. Here
that means surfaces of constant $z_3$. Their metric is given by 
\beq
ds_{\mbox{ind}}^2 = K_{a\bb} dz^a d\bz^{\bb}
\eeq 
Basically, we can say that the surfaces of constant axidilaton have a K\"ahler potential $K$ (for fixed $z_3$)
and the manifold where the $D7$-brane is wrapped a K\"ahler potential $\tK$ (for fixed $\eta$), where
$\tK$ is the Legendre transform of $K$ with respect to $x=\tau_2$. 

Now we have to find a solution to the equation for $\Phi$. To do this, we assume that $\Phi$ is a function
of $\rho=z_1\bz_1+z_2\bz_2$ which introduces an $SU(2)$ isometry in the metric. We get the equation
\beq
\det(\Phi_{a\bb}) = \pm \Phi^2 \ \ \ \ \Rightarrow \ \ \ \Phi_\rho^2+\rho\Phi_\rho\Phi_{\rho\rho}= \pm\Phi^2 .
\label{phieqn2}
\eeq
This equation is homogeneous in $\Phi$ which suggests a change of variable $\Phi=\exp(f)$ resulting in 
\beq
f'^2 + \rho f'(f''+f'^2)=\pm 1 ,
\label{feqn}
\eeq
we now define $y=f'$ and get a first order equation
\beq
\rho y y'= \pm1 -\rho y^3-y^2  .
\label{yeq}
\eeq
This type of equation is known as an Abel equation but unfortunately, to our knowledge, it does not have a solution  
in terms of elementary functions. It is easy however to find a solution as a series expansion or numerically.
Before doing that we rewrite the metric (\ref{m3v4}) to understand what choice of sign to do and which boundary
conditions to impose on $y(\rho)$.

Using that $\Phi$ is only a function of $\rho$, the metric (\ref{m3v4}) can be written as
\beq
ds^2 = X \left(\Phi_{\rho} dz^ad\bz^{\ba} + \Phi_{\rho\rho} \bz^a dz^a\, z^{\bb} d\bz^{\bb} \right)
 - \frac{1}{4} \Phi X_{\eta\eta} d\eta^2 - \frac{1}{4X_{\eta\eta}\Phi} 
\left(d\theta -iX_{\eta}\Phi_{\rho} ( \bz^adz^a-z^{\ba}dz^{\ba})\right)^2 .
\label{m3v5}
\eeq
It is now convenient to change into angular variables $\vartheta$, $\psi$, $\phi$: 
\beq
z_1 = \sqrt{\rho} \cos \frac{\vartheta}{2} e^{\frac{1}{2}i(\psi+\phi)},   \ \ \ \ 
z_2 = \sqrt{\rho} \sin \frac{\vartheta}{2} e^{\frac{1}{2}i(\psi-\phi)} ,
\label{ztophi}
\eeq
and introduce the 1-forms $\sigma_1$, $\sigma_2$, $\sigma_3$ through:
\beq
\sigma_1 + i \sigma_2 = e^{-i\psi} (d\vartheta + i\sin\vartheta d\phi)    ,\ \ \
\sigma_3 = d\psi+\cos\vartheta\, d\phi ,
\label{sigmas}
\eeq
which result in 
\beqa
|dz_1|^2 + |dz_2|^2 &=& \frac{1}{4\rho} d\rho^2 + \frac{\rho}{4} (\sigma_1^2+\sigma_2^2+\sigma_3^2) \nonumber , \\ 
|\bz_1dz_1+\bz_2dz_2|^2 &=& \frac{1}{4}d\rho^2 + \frac{1}{4}\rho^2 \sigma_3^2  \label{ztosigma} , \\
\bz^{\ba}dz^a-z^ad\bz^{\ba} &=& i\rho \sigma_3 \nonumber .
\eeqa
Finally, it is convenient to change coordinates from $\eta$ to $X$ using that
from equation (\ref{Xint}) we have:
\beq
          X_{\eta} = \left(A\pm\frac{X^3}{4}\right)^{\frac{1}{3}}, 
\ \ \ X_{\eta\eta} = \frac{X^2}{4} \left(A\pm\frac{X^3}{4}\right)^{-\frac{1}{3}} .
\eeq
With these results and eqns.(\ref{ztosigma}), the metric (\ref{m3v5}) can be written as
\beqa
ds^2 &=& \pm \frac{X}{y} e^f \frac{d\rho^2}{4\rho} + \frac{\rho X}{4} y e^f (\sigma_1^2+\sigma_2^2) \pm \frac{\rho X}{4y} e^f \sigma_3^2
\mp \frac{1}{16} e^f \frac{X^2}{(A\pm\frac{X^3}{4})} dX^2 \\
     &&  \mp \frac{(A\pm\frac{X^3}{4})^{\frac{1}{3}} }{X^2} 
             \left(d\theta + (A\pm\frac{X^3}{4})^{\frac{1}{3}} \rho y e^f \sigma_3\right)^2 .
\label{m3v6}
\eeqa
If we choose $X$ to be positive then, to get a positive definite metric we have to choose the upper signs everywhere. However, this 
has to be complemented by taking $A<0$ and $0<\frac{X^3}{4}<A$. Up to a rescaling, we can take $A=1/4$, $0<X<1$. It is also
convenient to introduce a new radial coordinate $u=\sqrt{\rho}$. 
The final form of the solution is
\beqa
ds^2 &=& X e^f \left[\frac{1}{y} du^2 + \frac{1}{4} u^2\left(y(\sigma_1^2+\sigma_2^2)+\frac{1}{y} \sigma_3^2\right)\right]   
+ \frac{1}{4} e^f \frac{X^2}{1-X^3} dX^2 \nonumber \\ 
     && + 4^{-\frac{1}{3}} e^{-f} \frac{(1-X^3)^{\frac{1}{3}}}{X^2}
        \left(d\theta - 4^{-\frac{1}{3}} u^2 y e^f (1-X^3)^{\frac{1}{3}} \sigma_3 \right)^2  ,\\
\phi &=& -\ln \left(N 4^{-\frac{1}{3}} e^f (1-X^3)^{\frac{1}{3}} \right) , \nonumber \\
\chi &=& N \theta , \nonumber 
\label{m3v7}
\eeqa
where we remind the reader that $y(\rho)$ is a function satisfying (\ref{yeq}) with a plus sign. Also, $y=\frac{df}{d\rho}$
and $\rho=u^2$. The values of $\phi$ and $\chi$ follow from the initial ansatz:
\beq
\tau = \chi + i e^{-\phi} = z_3 = \theta + i x = \theta -i \tK_{\eta} = \theta -i\Phi X_{\eta} 
      = \theta + i 4^{-\frac{1}{3}} e^f (1-X^3)^{\frac{1}{3}}  .
\label{taud4}
\eeq
In the metric (\ref{m3v7}), the manifold spanned by $(u,\vartheta,\psi,\phi)$ closes smoothly at $u=0$ only 
if $y(u=0) = 1$. This is because $d\Omega_{[3]}^2=(\sigma_1^2+\sigma_2^2+\sigma_3^2)/4$ is the metric of a 
round three sphere. In fact, the equation (\ref{yeq}) that $y$ satisfies actually implies that the only 
solution regular at $\rho=0$ satisfies precisely $y(0)=1$. We can get a series expansion as
\beq
y(\rho) = 1-\frac {1}{3}\rho+\frac {7}{36}{\rho}^{2}-\frac {16}{135}{\rho}^{3}+\frac {931}{12960}{\rho}^{4}
          -\frac {163}{3780}{\rho}^{5}+\ldots
\label{yseries}
\eeq
The asymptotic behavior follows from the same equation (\ref{yeq}) and is given by
\beq
y (\rho) \simeq \rho^{-\frac{1}{3}},\ \ \ \ \mbox{for}\ \ \rho\rightarrow\infty .
\label{yasympt}
\eeq
With this solution for $y(\rho)$ , the manifold over which the $D7$-brane is wrapped, namely the one
spanned by $(u,\vartheta,\psi,\phi)$ is non-singular and also non-compact. On the other hand, the full
metric is singular at $X=0$ and $X=1$. This is worse than the previous case where the 
metric was singular only at the 'core' where $\tau_2=0$ which here would be $X=1$. It is not clear to us the interpretation
of the extra singularity at $X=0$. 

It is instructive to perform a similar calculation as in this section for the case $p=1$. This also gives two
singularities in the radial direction as opposed to what we obtained in the previous section, namely only a singularity
at $x=0$. This indicates that perhaps an improvement of the method used in this section can lead to avoid
the singularity at $X=0$ but we did no explore that. For completeness, we write down the solution for $p=1$ that
we just mentioned:
\beqa
ds^2 &=& \sin u \cosh^{\frac{4}{3}}\!\chi\,d\vartheta^2 + \frac{4}{9}\sin u\cosh^{\frac{2}{3}}\!\chi\,(d\chi^2+du^2)+
  \frac{\cos^{\frac{2}{3}}\!u}{\cosh^{\frac{2}{3}}\!\chi\sin u} (d\theta+\sinh\chi\cos^{\frac{2}{3}}\!u\, d\vartheta)^2  ,
   \nonumber \\
\tau &=& \theta + i (\cosh\chi\cos u)^{\frac{2}{3}} ,
\eeqa
where the variables run over: $0<u<\frac{\pi}{2}$, $0<\chi<\infty$, $0<\vartheta<2\pi$, $0<\theta<2\pi$. The metric
is singular at $u=0$ and $u=\pi/2$. 

Finally let us mention that from the solution (\ref{m3v7}) we can get, using (\ref{cKmetric}), a 6-dimensional
Ricci flat metric. 

\section{Generalizations}

The solutions we have obtained up to now are generalizations of the ``circularly symmetric'' D7-brane of \cite{Bergshoeff:1996ui}.
These solutions have a curvature singularity where $\tau_2=0$. This follows from the equation for the 
metric (\ref{taueq}):
\beq
R_{\mu\nu} = \frac{1}{4\tau_2^2}\left(\partial_\mu\tau\partial_\nu\btau+\partial_\nu\tau\partial_\mu\btau\right) .
\eeq
Instead, in the original $D7$-brane construction \cite{Greene:1989ya, Gibbons:1995vg} a solution was proposed where
$\tau_2>0$ everywhere. That was accomplished by using the freedom in choosing the functions $\Omega$, $\tau$ in the solution
(\ref{D7metric}). One chooses a holomorphic function $\tau(z)=\tau_1+i\tau_2 $ that maps the complex plane onto the 
fundamental domain: $-\frac{1}{2}\le \tau_1 \le \frac{1}{2}$, $\tau_2>0$, $|\tau|\ge 1$.  
Such a function has cuts where $\tau$ jumps by an \sltz\ transformation. These transformations now include, 
besides $\tau\rightarrow \tau+1$ also $\tau \rightarrow -1/\tau$ and combinations. In that case $\tau_2(z,\bz)$ also has cuts 
and cannot enter directly in the metric. However, choosing $\Omega$ adequately one can make the combination 
$\Omega(z)\, \bar{\Omega(\bz)}\, \tau_2(z,\bz)$ that appears in the metric invariant under modular transformations and therefore,
a function of $z$, $\bz$ with no cuts. In spite of the fact that $\tau_2>0$ everywhere, 
the resulting solution still has curvature singularities because $\partial_\mu\tau$  diverges at points where $\tau=i$ or
$\tau=e^{i\pi/3}$. This singularities are milder since at a distance $\delta$ from the singularity 
$\partial_\mu\tau \sim \delta^{-\alpha}$, $\alpha<1$ and therefore the action obtained by integrating the Ricci 
scalar on a transverse plane is finite.

 In this section we generalize this type of solution to a $D7$ brane wrapped on a 2 dimensional manifold using the same
calculations as in section \ref{d=2}. We are looking then for a function 
$K(z_1,z_2,\bz_1,\bz_2)$ such that 
\beq
K_{1\bo} K_{2\bt} - K_{1\bt} K_{2\bo} = |\Omega|^2 \tau_2 .
\label{Keq2b}
\eeq
Now assume that $K$ is only a function of the real part of $z_1$ and introduce real coordinates through
\beq
z_1 = y+i \phi .
\eeq
Furthermore we then assume that $\tau$ and $\Omega$ are functions only of $z_2$. Thus,  eq.(\ref{Keq2b}) becomes
\beq
K_{yy} K_{2\bt} - K_{y 2} K_{y\bt} = 4 |\Omega(z_2)|^2 \frac{\tau(z_2)-\btau(\bz_2)}{2i} .
\eeq
The function $\tau(z_2)$ can have cuts but $\Omega$ is then chosen such that the right hand side
of the equation is well defined throughout the $z_2$ plane. 
 Now, we use again a Legendre transformation
\beq
\tK(\eta,z_2,\bz_2) = K(y,z_2,\bz_2) - \eta y, \ \ \eta=K_y ,
\eeq
which results in 
\beq
y=-\tK_{\eta}, \ \  K_{2 y} = -\frac{\tK_{2\eta}}{\tK_{\eta\eta}}, 
\ \ K_{2\bt}=\tK_{2\bt} -\frac{\tK_{2\eta}\tK_{\bt\eta}}{\tK_{\eta\eta}},
\ \ \ K_{yy} = - \frac{1}{\tK_{\eta\eta}} .
\eeq
Using this we linearize eq.(\ref{Keq2b}):
\beq
\tK_{2\bt} + 4\, \Omega\,\bar{\Omega}\,\tau_2\, \tK_{\eta\eta}  = 0 .
\label{Klin2}
\eeq
The assumption that $\Omega$ is independent of $y$ is important since $y=-\tK_{\eta}$ and any $y$ dependence
will render the equation non-linear. The metric is
\beqa
ds^2 &=& \tK_{2\bt} dz_2 d\bz_2 - \frac{1}{4\tK_{\eta\eta}} \left| dz_1 + 2 \tK_{2\eta} dz_2 \right|^2 \\
 &=& \tK_{2\bt} dz_2d\bz_2 - \frac{\tK_{\eta\eta}}{4} d\eta^2 
     - \frac{1}{4\tK_{\eta\eta}} \left(d\phi - i (\tK_{2\eta}dz_2-\tK_{\bt\eta}d\bz_2)\right)^2  .
\eeqa
We still have to solve (\ref{Klin2}). Using separation of variables we can write a generic solution as
\beq
\tK = \int_{-\infty}^{+\infty} dk\, e^{ik\eta}\, \Phi_k(z_2,\bz_2) ,
\eeq
where
\beq
\partial_2\bpartial_{\bt} \Phi_k = 4\, k^2\, \Omega\,\bar{\Omega}\,\tau_2\, \Phi_k .
\label{phik}
\eeq
If we use the same $\Omega$ as in \cite{Greene:1989ya}, this becomes
\beq
\partial_2\bpartial_{\bt} \Phi_k = 4 k^2\, \eta^2(z_2)\, \bar{\eta}^2(\bz_2)\, \tau_2\, \Phi_k ,
\eeq
where $\eta(z)$ is Dedekinds $\eta$ function:
\beq
\eta = q^{1/24} \prod_n(1-q^n) ,
\eeq
with $q=e^{2\pi i\tau(z_2)}$ and $\tau(z_2)$ is defined implicitly through the equation
\beq
j(\tau) = z_2 ,
\eeq
where $j$ can be written in terms of Jacobi $\theta$-functions as:
\beq
j(\tau) = \frac{(\theta_2^8(\tau)+\theta_3^8(\tau)+\theta_4^8(\tau))^3}{\eta^{24}(\tau)} ,
\eeq
and has the virtue of being modular invariant.
 With these values of $\Omega$ and $\tau$ we should solve eq.(\ref{phik}):
\beq
\partial_{2\bt} \Phi_k = 4\, k^2\, \eta^2\,\bar{\eta}^2\,\tau_2\, \Phi_k .
\eeq
Unfortunately we were not able to find a solution to this equation so we leave the metric as it is in terms
of the function $\Phi_k$ that should be determined, perhaps numerically.

Another possible generalization which we did not consider is to introduce D3 branes parallel to the 
D7 branes along the directions that we did not wrap. It seems possible that one can find such
solutions along the lines of \cite{Kehagias:1998gn}.

\section{Conclusions}
\label{5sec}

We described a generalization of the D7 brane solution such that the world volume of the brane
is $\bbr{[1,5]}\times T^2$ where the metric on $T^2$ is an arbitrary metric with translation invariance
along one of the directions of the torus. The solution has a singularity at a finite radius in the same way
as the flat brane has. Also, the function $\tau$ has a cut with a jump of the form $\tau\rightarrow \tau+N$
with $N$ an integer counting the number of branes. 
In analogy with the flat brane case, we further discussed how we can make this singularity milder by introducing 
additional cuts in the function $\tau$ where $\tau$ is identified up to an \sltz\ transformation including 
$\tau\rightarrow -1/\tau$. In this case we did not write the metric explicitly but left it expressed in terms 
of the solutions of a certain linear equation.

All these supergravity backgrounds were obtained by using the same Legendre transform method which is sometimes used to find 
4-dimensional hyperK\"ahler metrics. 

We also considered the case where we wrapped the brane on a 4-dimensional manifold. In this case the equation cannot
be simplified as much as in the previous case and we find only a solution that is singular close and far from
the brane. The interpretation of this solution is not clear. 

We would like to point out that the solution with the D7-brane wrapped on $T^2$ depends on an arbitrary function
that determines the metric on $T^2$. Recently, it has been suggested that it might be possible to obtain solutions 
for black p-branes where the horizon is not uniform \cite{Gubser:2001ac}. In our case there is no horizon but a 
singularity. However, it would be interesting to see if by applying T and S-dualities to our solution one can shed 
some light on the problem of non uniform branes.

Another point that would be interesting to pursue is the inclusion of D3 branes in this set up. This would lead
to field theories with fields in the fundamental of the type discussed in detail in \cite{Kruczenski:2003be}. By
immersing the D7 branes in \ads{5} the gauge group on the D7 branes becomes a global (flavor) group in the
boundary theory. 

Finally, it would also be interesting to understand the relation, if any, to work done in the case of 
$M5$-branes \cite{Fayyazuddin:1999zu,Maldacena:2000mw}. Those solutions were mostly for branes wrapped on 
manifolds of constant curvature but could be generalized to a situation similar to the one analyzed here.

\section{Acknowledgments}
This work was supported in part by NSF through grant PHY-0331516. In addition, I would like to thank 
the Perimeter Institute for Theoretical Physics for hospitality and partial support while this work 
was being done. I am also grateful to Carlos Nu\~nez for comments on a draft of the paper.

\section*{Appendix A}

In this appendix we solve the equation
\beq
\tK_{xx}+ x \tK_{\eta\eta}= 0 .
\label{AKeq}
\eeq
 The equation has similar properties to the Laplace equation. The variable's domain of variation 
is: $-\infty<\eta<\infty$, $0\le x<\infty$. We are going to use as boundary conditions
\beq
\left.\tK(x,\eta)\right|_{x=0} = h(\eta) ,
\label{Abc}
\eeq
and that $\tK$ remains finite for $x\rightarrow\infty$. Since the equation is linear, it 
 can be trivially solved by separation of variables with the result
\beq
\tK = - 3^{\frac{2}{3}} \Gamma\left(\frac{2}{3}\right) \int_{-\infty}^{+\infty}dk\, h(k) e^{ik\eta} \Ai\left(k^{\frac{2}{3}} x\right)  ,
\label{Asolution}
\eeq
where $\Ai$ is the Airy function\footnote{In this appendix we use several properties of special functions which 
can be obtained from \cite{Gradshteyn} or by using a computer algebra program such as Maple v.9} which solves the equation
\beq
\frac{d^2\Ai(x)}{dx^2} - x \Ai(x) = 0 ,
\label{AAiry}
\eeq
and vanishes at infinity. It can be written in terms of the Bessel function $\besselK_{\frac{1}{3}}$ as:
\beq
\Ai(x) = \frac{1}{\pi} \sqrt{\frac{x}{3}} \besselK_{\frac{1}{3}}\!\left(\frac{2}{3}x^{\frac{3}{2}}\right) .
\eeq
There is an independent solution that blows up as $x\rightarrow\infty$ and therefore
should be discarded since we want $\tK$ to be finite. In $\tK$ we included a (negative) constant coefficient in front 
because the function $\Ai$ is normalized such that
\beq
\Ai(0) = \frac{1}{3^{\frac{2}{3}}\Gamma\left(\frac{2}{3}\right)} .
\label{AAiry2}
\eeq
This means that
\beq
\tK(x=0,\eta) = - h(\eta) = - \int_{-\infty}^{+\infty} h(k) e^{ik\eta}  ,
\label{Abc2}
\eeq
namely $h(k)$ is the Fourier transform of $h(\eta)$. Inverting this we arrive at
\beq
\tK = - 3^{\frac{2}{3}} \Gamma\left(\frac{2}{3}\right) \int_{-\infty}^{+\infty} dk\,
 \frac{1}{2\pi} \int_{-\infty}^{+\infty}d\teta\, e^{-ik\teta} h(\teta) e^{ik\eta} \Ai\left(k^{\frac{2}{3}} x\right) .
\label{AKinter}
\eeq
Interchanging the order of integration and performing the $k$ integral we obtain
\beq
\tK(\eta,x) = a_0(x^3-3\eta^2) - \frac{3^{\frac{1}{3}}}{2\pi^2} \Gamma\left(\frac{2}{3}\right)^3 
              \int_{-\infty}^{+\infty} 
              \frac{x}{\left[(\teta-\eta)^2+\frac{4}{9}x^3\right]^{\frac{5}{6}}} 
              h(\teta)d\teta , 
\label{AKsol}
\eeq
where, for completeness, we added a trivial solution $a_0(x^3-3\eta^2)$ to what we obtained from (\ref{AKinter}).
The last expression is used in the main text. 

\section*{Appendix B}

 In this appendix we summarize some well known properties of K\"ahler manifolds which are useful in 
the main text. On a K\"ahler manifold, the metric can be written through a K\"ahler potential as
\beq
ds^2 = \partial_{a\bb} K dz^a d\bz^{\bb} .
\eeq
 The non-vanishing components of the Levi-Civita connection can be computed using the usual definition and are given by 
\beq
\Gamma^c_{ab} = g^{c\bd}\partial_{ab\bd} K ,\ \ \ \Gamma^{\bc}_{\ba\bb} = g^{\bc d}\partial_{\ba\bb d} K  .
\eeq
From here, we obtain that the only non-vanishing components of the Ricci tensor are
\beq
R_{a\bb} = - \partial_{\bb} \Gamma^{c}_{ac} = -\partial_{\bb} ( g^{c\bd}\partial_a g_{c\bd} )
         = - \partial_{a\bb} \ln\det g_{c\bd} .
\eeq
The determinant in the last expression is that of the matrix $g_{a\bb}$ as indicated. Note that the
determinant of the metric is: $\det g = (\det g_{a\bb})^2$.
One can introduce a vielbein: $e^A_a$, $\bar{e}^{\bA}_{\ba}$ such that
\beq
g_{a\bb} = \partial_{a\bb} K = e^{C}_a \bar{e}^{\bC}_{\bb} .
\eeq
Deriving the last equality with respect to $z^d$ and antisymmetrizing in $a,d$ we obtain the relation
\beq
\partial_{d} e^C_a \bar{e}^{\bC}_{\bb} - \partial_{a} e^C_d \bar{e}^{\bC}_{\bb} =
e^C_d \partial_{a} \bar{e}^{\bC}_{\bb} - e^C_a \partial_{d} \bar{e}^{\bC}_{\bb}  .
\eeq
Using this result in the definition of spin connection 
\beq
\omega^{MNP} =\half e^{nN} e^{mM} e^{P}_{[n,m]} - \half e^{nP} e^{mM} e^{N}_{[n,m]}-\half e^{mN} e^{nP} e^{M}_{[n,m]} ,
\eeq
we get that the only non-vanishing components of $\omega$ are
\beq
\omega^{CA\bB} = - \omega^{C\bB A} = - e^{a\bB} e^{\bb C} \partial_{\bb} e^{A}_a ,
\eeq
and their complex conjugates. From here, we find that the $U(1)$ part of the connection is given by
\beq
\sum_A \omega_{\ba}{}^A{}_A = \sum_A \omega_{\ba \bA A} = -e^b_A \partial_{\ba} e^A_b = -\partial_{\ba} \ln\det e^A_a .
\eeq
Notice also that 
\beq
\det g_{a\bb} = \det e^A_a \det \bar{e}^{\bB}_b = |\det e^{A}_a|^2 .
\eeq


\begin{thebibliography}{99}        

\bibitem{Polchinski:1995mt}
J.~Polchinski,
``Dirichlet-Branes and Ramond-Ramond Charges,''
Phys.\ Rev.\ Lett.\  {\bf 75}, 4724 (1995)
[arXiv:hep-th/9510017].

\bibitem{Giveon:1998sr}
A.~Giveon and D.~Kutasov,
``Brane dynamics and gauge theory,''
Rev.\ Mod.\ Phys.\  {\bf 71}, 983 (1999)
[arXiv:hep-th/9802067].

\bibitem{Katz:1996th}
S.~Katz and C.~Vafa,
``Geometric engineering of N = 1 quantum field theories,''
Nucl.\ Phys.\ B {\bf 497}, 196 (1997)
[arXiv:hep-th/9611090].

\bibitem{Vafa:1996xn}
C.~Vafa,
``Evidence for F-Theory,''
Nucl.\ Phys.\ B {\bf 469}, 403 (1996)
[arXiv:hep-th/9602022].

\bibitem{Greene:1989ya}
B.~R.~Greene, A.~D.~Shapere, C.~Vafa and S.~T.~Yau,
``Stringy Cosmic Strings And Noncompact Calabi-Yau Manifolds,''
Nucl.\ Phys.\ B {\bf 337}, 1 (1990).

\bibitem{Gibbons:1995vg}
G.~W.~Gibbons, M.~B.~Green and M.~J.~Perry,
``Instantons and Seven-Branes in Type IIB Superstring Theory,''
Phys.\ Lett.\ B {\bf 370}, 37 (1996)
[arXiv:hep-th/9511080].

\bibitem{Bergshoeff:1996ui}
E.~Bergshoeff, M.~de Roo, M.~B.~Green, G.~Papadopoulos and P.~K.~Townsend,
``Duality of Type II 7-branes and 8-branes,''
Nucl.\ Phys.\ B {\bf 470}, 113 (1996)
[arXiv:hep-th/9601150].

\bibitem{Einhorn:2000ct}
M.~B.~Einhorn and L.~A.~Pando Zayas,
``On seven-brane and instanton solutions of type IIB,''
Nucl.\ Phys.\ B {\bf 582}, 216 (2000)
[arXiv:hep-th/0003072].

\bibitem{Bergshoeff:2002mb}
E.~Bergshoeff, U.~Gran and D.~Roest,
``Type IIB seven-brane solutions from nine-dimensional domain walls,''
Class.\ Quant.\ Grav.\  {\bf 19}, 4207 (2002)
[arXiv:hep-th/0203202].

\bibitem{Schwarz:qr}
J.~H.~Schwarz,
``Covariant Field Equations Of Chiral N=2 D = 10 Supergravity,''
Nucl.\ Phys.\ B {\bf 226}, 269 (1983).

\bibitem{Lindstrom:rt}
U.~Lindstrom and M.~Rocek,
``Scalar Tensor Duality And N=1, N=2 Nonlinear Sigma Models,''
Nucl.\ Phys.\ B {\bf 222}, 285 (1983).

\bibitem{Kehagias:1998gn}
A.~Kehagias,
``New type IIB vacua and their F-theory interpretation,''
Phys.\ Lett.\ B {\bf 435}, 337 (1998)
[arXiv:hep-th/9805131].

\bibitem{Gubser:2001ac}
S.~S.~Gubser,
``On non-uniform black branes,''
Class.\ Quant.\ Grav.\  {\bf 19}, 4825 (2002)
[arXiv:hep-th/0110193].

\bibitem{Kruczenski:2003be}
M.~Kruczenski, D.~Mateos, R.~C.~Myers and D.~J.~Winters,
``Meson spectroscopy in AdS/CFT with flavour,''
JHEP {\bf 0307}, 049 (2003)
[arXiv:hep-th/0304032].

\bibitem{Fayyazuddin:1999zu}
A.~Fayyazuddin and D.~J.~Smith,
``Localized intersections of M5-branes and four-dimensional  superconformal field theories,''
JHEP {\bf 9904}, 030 (1999)
[arXiv:hep-th/9902210].

\bibitem{Maldacena:2000mw}
J.~M.~Maldacena and C.~Nunez,
``Supergravity description of field theories on curved manifolds and a no  go theorem,''
Int.\ J.\ Mod.\ Phys.\ A {\bf 16}, 822 (2001)
[arXiv:hep-th/0007018].

\bibitem{Gradshteyn}
I.S. Gradshteyn, I.M. Ryzhik, ``Table of Integrals Series and Products'', Sixth edition,
Academic Press (2000), San Diego, CA, USA, London, UK.


\end{thebibliography}
\end{document}